%% file: main.tex
\documentclass[9pt,conference]{IEEEtran}
\usepackage{cite}
\usepackage{amsmath,amssymb,amsfonts}
\usepackage{algorithmic}
\usepackage{graphicx}
\usepackage{textcomp}
\usepackage[hyphens]{url}
\usepackage{fancyhdr}
\usepackage{hyperref}
\usepackage{amssymb}
\usepackage{wasysym}
\usepackage{pifont}
\usepackage{caption}
\usepackage{subcaption}
\usepackage{mdframed}
\usepackage{array}
\usepackage{multirow}
\usepackage{makecell}
\usepackage{tabularx}
\usepackage{booktabs}
\usepackage{newtxtext}
\usepackage[dvipsnames]{xcolor}
\newcommand{\greencheck}{{\textcolor{ForestGreen}{\ding{51}}}}
\newcommand{\redcross}{{\textcolor{BrickRed}{\ding{55}}}}
\usepackage{array}
\usepackage{balance}
\usepackage{soul}
\newcommand{\name}{\textsc{CXL-Interference}}
\newcommand{\sota}{{state-of-the-art}}

\usepackage{tikz}
\usetikzlibrary{fit}

\begin{document}

\title{\name{}: Analysis and Characterization in Modern Computer Systems}

\author{
\IEEEauthorblockN{Shunyu Mao\textsuperscript{*}}
\IEEEauthorblockA{\textit{Tsinghua University}}
\and
\IEEEauthorblockN{Jiajun Luo\textsuperscript{*}}
\IEEEauthorblockA{\textit{Tsinghua University}}
\and
\IEEEauthorblockN{Yixin Li}
\IEEEauthorblockA{\textit{Tsinghua University}}
\and
\IEEEauthorblockN{Jiapeng Zhou}
\IEEEauthorblockA{\textit{Institute of Computing Technology} \\
\textit{Chinese Academy of Sciences}
}
\and
\IEEEauthorblockN{Weidong Zhang}
\IEEEauthorblockA{\textit{Alibaba Group}}
\and
\IEEEauthorblockN{Zheng Liu}
\IEEEauthorblockA{\textit{Alibaba Group \& Zhejiang University}}
\and
\IEEEauthorblockN{Teng Ma}
\IEEEauthorblockA{\textit{Alibaba Group}}
\and
\IEEEauthorblockN{Shuwen Deng\textsuperscript{\textdagger}}
\IEEEauthorblockA{\textit{Tsinghua University}}
}

\maketitle
\renewcommand{\thefootnote}{\fnsymbol{footnote}}
\footnotetext[1]{Co-first author.}
\footnotetext[2]{Corresponding author.}
\renewcommand{\thefootnote}{\arabic{footnote}}

\input{00_abstract}

\input{01_introduction}

\input{02_background}

\input{03_overview_setup}

\input{04_microbenchmarks}

\input{05_real_applications}
\input{06_hardware_influence_on_qos}

\input{05p_practical_suggestion}
\input{08_conclusion}

\bibliographystyle{IEEEtranS}
\bibliography{ref}

\end{document}

%% file: 00_abstract.tex
\begin{abstract}

Compute Express Link (CXL) is a promising technology that addresses memory and storage challenges.
Despite its advantages, CXL faces performance threats from external interference when co-existing with current memory and storage systems. This interference is under-explored in existing research. To address this, we develop CXL-Interplay, systematically characterizing and analyzing interference from memory and storage systems.
To the best of our knowledge, we are the first to characterize CXL interference on real CXL hardware. 
We also provide reverse-reasoning analysis with performance counters and kernel functions. In the end, we  
propose and evaluate mitigating solutions.

\end{abstract}

\maketitle
\pagestyle{plain}

%% file: 01_introduction.tex
 \section{Introduction \& Related Work}
\label{sec:introduction}

Compute Express Link (CXL)~\cite{compute2020compute} provides a comprehensive solution to existing memory wall problems~\cite{gholami2024ai} and has garnered significant attention from both industry and academia. CXL offers several advantages, including high bandwidth density and a standardized interface for memory expansion and pooling~\cite{cxl_optimization_sk}. These features make it an ideal solution to address the limitations of current cloud storage architectures, and have the potential to revolutionize data center architectures~\cite{bai2022analysis}. 
In industry, leading companies are actively exploring and adopting CXL to improve the efficiency and scalability of their data center products, including Intel~\cite{intelcxl}, Samsung~\cite{sumsungcxl}, and SK Hynix~\cite{skhynixcxl}.
Academically, researchers are investigating various aspects of CXL, including its architecture~\cite{scaleout_architecture}, performance implications~\cite{sun2023demystifying, tang2024exploring}, and potential security concerns~\cite{security_cxl_gpu}.

However, coexisting with current memory and storage systems in the server architecture, CXL faces understudied performance threats from external interference, which are not adequately reflected in existing research. 
Zooming out from CXL itself and considering its operational environment, CXL can potentially be affected by interference from 1) interactions between Main Memory (\textbf{MMEM}) and CXL and 2) interactions between neighboring storage components and CXL.
Avoiding interference and maintaining performance isolation are critical for performance sensitive applications.
For instance, MT$^{2}$~\cite{yi2022mt} tries to explore the interference between persistent memory (PMEM) and DRAM, and it identifies noisy neighbors among concurrent applications to mitigate memory traffic interference.
To the best of our knowledge, this \textbf{\textit{interference issue for CXL has not been extensively studied in existing research}}. Current CXL simulators typically introduce delay factors manually to simulate memory delays~\cite{yang2023cxlmemsim,puri2024dracksim}, which neither reflects the real-world environment nor considers interference. Moreover, existing hardware profiling studies on CXL primarily focus on evaluating the behavior of CXL under different operations and applications~\cite{sun2023demystifying,tang2024exploring,liu2024exploringevaluatingrealworldcxl}, without considering the impact of external environmental factors.

Given that, we develop \name{}, which \textbf{\textit{systematically characterizes and analyzes the potential reasoning of interference between memory/storage systems and CXL memory}} using carefully designed microbenchmarks and tailored real applications. We also discuss potential effective regulation that can be applied to fit real CXL interference scenarios.

First, in order to clearly identify and characterize CXL interference with memory and storage system, we develop a configurable microbenchmark and systematically go through the basic interference condition of CXL with MMEM and SSD system using \textit{two different real CXL hardware setups}.
As shown in Table{~\ref{tab:motivation}}, we are the first to demonstrate and characterize the {CXL interference using real CXL devices. }

Furthermore, we conduct reverse-reasoning based on profiling evaluation results, exploring kernel functions and hardware performance counters, including the new CXL-related counters.

For the real testing setup, we explore file system, database, and machine learning applications for storage-related testing environment. For the memory-related testing environment, we focus on large-language model (LLM), in-memory database, and graph computing applications.

Given the interplay results, we further explore software and discuss hardware intervening solutions to provide effective regulation for the system to mitigate the interference impact. 
Based on our evaluation, it is feasible to restore bandwidth of the MMEM by 204GB per second, which reaches 99\% of its original level, using software solutions, at the cost of a 5GB per second reduction in CXL bandwidth.

To summarize, this work makes the following contributions:
\begin{enumerate}
    \item Systematically characterize the interference issue
    in current server architectures among CXL, MMEM, and storage devices. To the best of our knowledge, this is the first study examining interference of CXL on real hardware.
    \item Conduct comprehensive microbenchmarks, evaluate memory-related and storage-related real applications and perform reverse-reasoning analysis for evaluation results.
    \item Explore and evaluate software-based intervention solutions to address the interference problem and mitigate its impact on real hardware.
    
\end{enumerate}

\begin{table}[t]
    \caption{Comparison of \name{} with \sota{} CXL characterization work. 
    To the best of our knowledge, we are the first to investigate interference and regulation of real CXL devices and conduct reverse-reasoning analysis.
  }
    \centering
    \renewcommand{\arraystretch}{1.3}
    \begin{tabular}{cccccc}
    \hline
         &\thead{\textbf{Real ASIC}\\\textbf{Device}} & \thead{\textbf{Inter-}\\\textbf{ference}} & \thead{\textbf{Regulation}} & \thead{\textbf{Reverse-}\\\textbf{reasoning}\\\textbf{Analysis}}  \\ \hline
         \thead{Emulation works\\ \cite{lee2023memtis,song2023lightweight,yang2023cxlmemsim}}  & \redcross &  \redcross & \redcross & \redcross  \\ \hline
        Sun~\cite{sun2023demystifying} & \greencheck  & \redcross & \redcross & \redcross  \\ \hline
        Tang~\cite{tang2024exploring} & \greencheck  & \redcross & \redcross & \redcross \\ \hline
        \thead{ Liu and \\ Wang~\cite{liu2024exploringevaluatingrealworldcxl} }& \greencheck  & \redcross & \redcross & \redcross\\ \hline
        \textbf{Our Work} & \greencheck & \greencheck & \greencheck & \greencheck  \\ \hline
        
    \end{tabular}
    \label{tab:motivation}
\end{table}

The
code used in this paper will be released under open-source
license at https://code-to-be-announced.

%% file: 02_background.tex
\section{Background \& \name{} Setup}
\label{sec:background}

\subsection{Compute Express Link (CXL)}

Compute Express Link (CXL)~\cite{compute2020compute}, first developed in 2019, is an open standard interconnect designed to enhance the performance of data-centric applications by enabling high-speed, low-latency communication between CPU/accelerators, memory, and storage.

The CXL protocol stack is divided into three parts: CXL.io, CXL.cache, and CXL.mem, each serving distinct purposes to facilitate various types of data transmission and memory access. CXL.io ensures compatibility with PCIe, enabling traditional I/O operations and device discovery. CXL.cache supports accelerators accessing CPU's internal caches and host memory. CXL.mem allows direct memory access and pooling, extending memory across multiple devices.

CXL devices are classified into three types, each tailored for specific applications. Type-1 devices, such as SmartNICs, utilize CXL.io and CXL.cache to facilitate communication with host DDR memory. Type-2 devices, including GPUs, ASICs, and FPGAs, use CXL.io, CXL.cache, and CXL.mem to share memory resources with the processor. Type-3 devices rely on CXL.io and CXL.mem to enable memory expansion and pooling. Type-3 devices enable extra DRAM capacity and bandwidth.

There are currently two types of CXL devices: FPGA-based and ASIC-based.  Intel Agilex FPGA integrates CXL hard IP, which is a standard component in recent CXL studies (e.g., Sun et al~\cite{sun2023demystifying}, Zhong et al~\cite{zhong2024managing}). Besides, emerging ASIC-based CXL memory devices are becoming available, with promises from several vendors such as Samsung~\cite{sumsungcxl}, Montage~\cite{montagecxl} and Micron~\cite{microcxl}.

\subsection{Memory System and CPUless Node}

DRAM is widely used in memory systems as the main memory. It is typically interconnected via a parallel interface. 
However, the limited capacity and bandwidth blocks its scalability for memory-intensive applications.

Secondary memory such as persistent memory (PMEM) can meet the demands of memory-intensive applications, and its large capacity makes it an alternative to DIMM. PMEM can be identified as the memory of an additional node without CPU. 

As a type of secondary memory, CXL memory provides CPUless node interface. However, unlike PMEM, CXL devices rely on PCIe, utilizing its physical layer and provides multiplexing features, resulting in physical topologies akin to those of PCIe.

%% file: 03_overview_setup.tex
\subsection{\name{} Hardware Setup}

In this work, we systematically characterize and explore how CXL memory interferes with other memory-related access, specifically MMEM access, and storage-related access such as NVMe or SAS SSD, using microbenchmarks and real applications. 

Table~\ref{tab:exp_setup} demonstrates the experimental setup of \name{}. We have two systems that will be referred to later as System A and System B respectively. System A has SAS SSD and an ASIC-based CXL memory device manufactured by Montage. System B consists of NVMe SSD and an Intel FPGA-based CXL memory device. 

\begin{figure}
    \centering
    \includegraphics[width=0.95\linewidth,trim={0 0 0 130},clip]{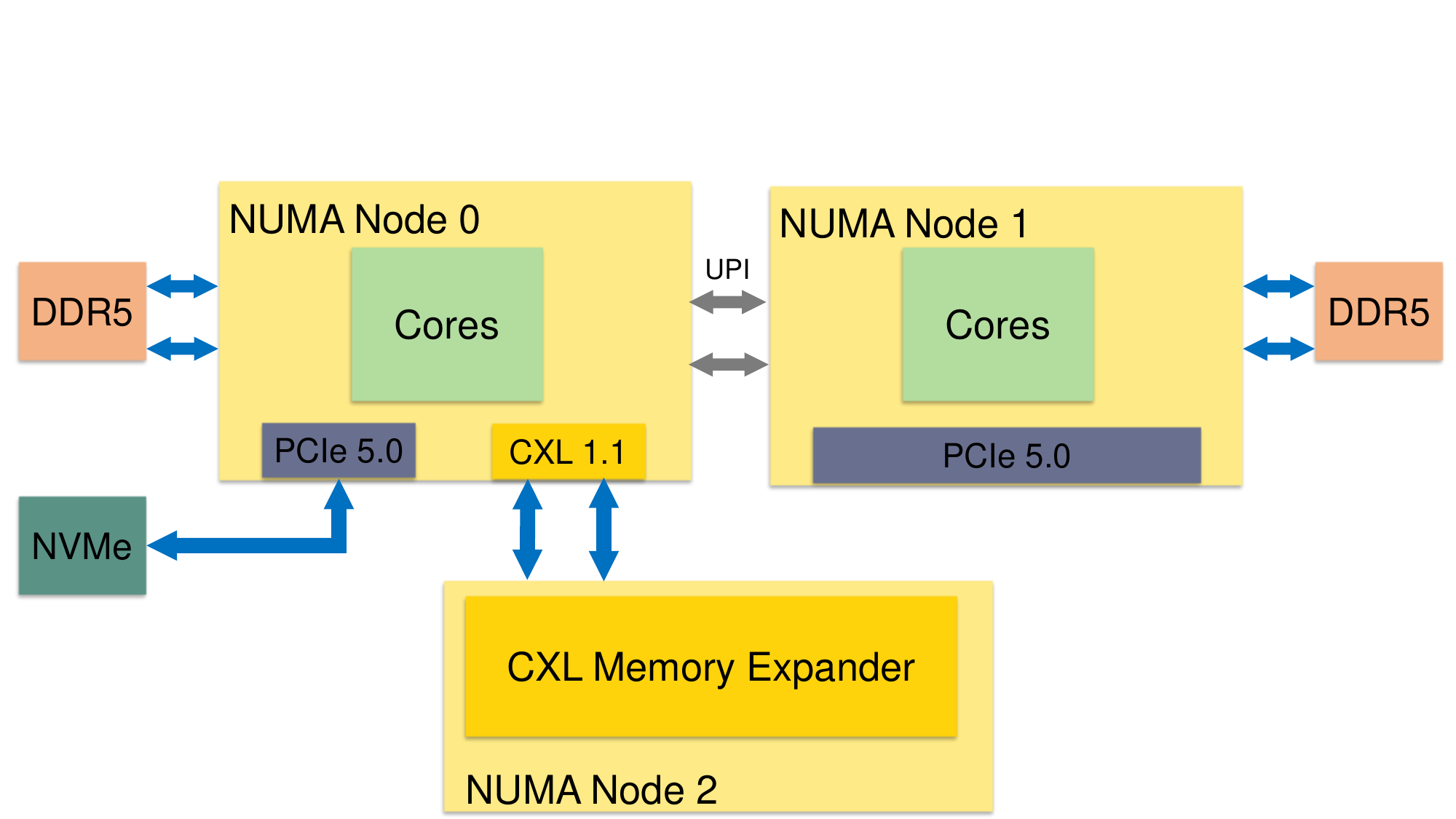}
    \caption{  {Schematic illustration of hardware setup}}
    \label{fig:setup}
\end{figure}

  {
Figure{~\ref{fig:setup}} illustrates the topology of our hardware setup, where NUMA node 0 and NUMA node 1 each possess their own cores and DDR5 DRAM, while the CXL Memory Expander is situated on NUMA node 2 with zero processor core, directly connected to Node 0. Therefore, tasks are always allocated to the cores on Node 0.
}
\begin{table}[t]
    \caption{Experimental setup of \name{}.}
    \centering
    \renewcommand{\arraystretch}{1.3}
    \begin{tabular}{ccc}
        \hline & \textbf{System A} & \textbf{System B} \\ \hline
        CPU & \thead{4th Gen Xeon Platinum \\ 4.0 GHz} & \thead{4th Gen Xeon Gold \\ 4.0 GHz} \\ \hline
        \thead{Number of  physical\\ cores  per socket} & 48 & 32 \\ \hline
        \thead{Memory per socket \\ Channels  Speed} & \thead{1TB,   {8 Channel}, \\ DDR5-4800} & \thead{256GB,   {8 Channel}, \\ DDR5-4800} \\ \hline
        Storage  & SAS SSD & \thead{NVMe PCIe 4.0 \\ MLC SSD}  \\ \hline
        File System & XFS  & XFS \\ \hline
        CXL memory device & \thead{Montage ASIC \\ 64GB} & \thead{Intel Agilex FPGA \\ 32GB DDR4-2400} \\ \hline
        OS & CentOS 7 & Ubuntu 22.04.4 LTS \\ \hline
    \end{tabular}
    \label{tab:exp_setup}
\end{table}

%% file: 04_microbenchmarks.tex
\section{\name{} Microbenchmarks}
\label{sec:microbenchmarks}

In this section, we set up comprehensive microbenchmarks to examine the CXL interference problem with SSD and MMEM.

\subsection{Microbenchmark Setup}

In our microbenchmark, three different memory-related operations (load (\textit{ld}), store (\textit{st}), and non-temporal store (\textit{ntst}))~\footnote{We do not have non-temporal load mainly considering its sparse applicability, which are demonstrated in a lot of works~\cite{intelsdm,ramesh2024hint,ex86}} 
and two different storage-related operations (random-read (\textit{randread}) and random-write (\textit{randwrite})) are cross-evaluated.

For memory-related operations,
we develop our microbenchmark by measuring the bandwidth of CXL memory and MMEM with cxlMemTest~\cite{sun2023demystifying}.
We carry out our experiments with all possible number of threads. For better measurement, we disable Hyperthreading~\cite{smt}, lock CPU's frequency to $2$GHz, and clear cache before each test. We also allocate main process and interfering process to separate cores of the same NUMA node. The measurement is repeated multiple times and get average. We also carry out case studies on random/sequential storage access pattern and \texttt{MOVDIR64B} instruction.

For storage-related operations, 
we employ {FIO}~\cite{fio} with \textit{libaio} as \textit{ioengine} to perform sustained \textit{randread}/\textit{randwrite} operations with an \textit{iodepth} of $16$ and a \textit{block size} of $4$KB for $20$ seconds.

\begin{table*}[t!]
    \centering
    \caption{The schematic illustration of bandwidth across different systems and configurations. In each box, the x-axis and y-axis represent the operation of the background and the device under test, respectively. We categorize the influence into seven types, each represented by a color in the colorbar: ``Deep Suppression", below -40\%; ``Moderate Suppression", -40\% to -10\%; ``Mild Suppression", -5\% to -10\%; ``No Influence", within -5\%; Promotion follows the same pattern. 
    }

    \begin{tabular}{|>{\centering\arraybackslash}m{0.07\textwidth}|>{\centering\arraybackslash}m{0.07\textwidth}|>{\centering\arraybackslash}m{0.17\textwidth}|>{\centering\arraybackslash}m{0.17\textwidth}|>{\centering\arraybackslash}m{0.17\textwidth}|>{\centering\arraybackslash}m{0.17\textwidth}|}
        \hline
        \multicolumn{2}{|c|}{Evaluated Device} & MMEM & SSD & \multicolumn{2}{c|}{CXL} \\ \hline
        \multicolumn{2}{|c|}{Background} & \multicolumn{2}{c|}{CXL} & MMEM & SSD \\ \hline
        \multirow{9}{*}{Bandwidth} & System A & \vspace{5pt}\includegraphics[width=0.95\linewidth,trim={0 0 0 0},clip]{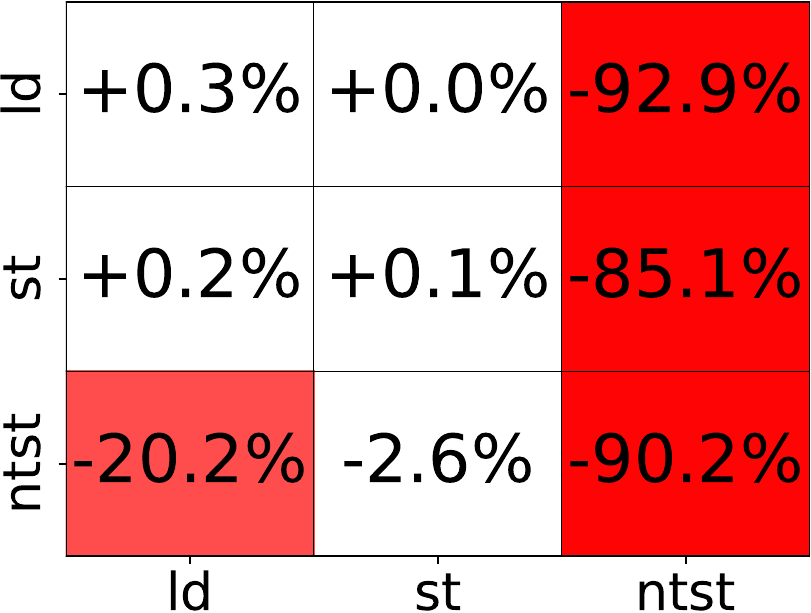} &
        \vspace{5pt}\includegraphics[width=0.95\linewidth,trim={0 0 0 90},clip]{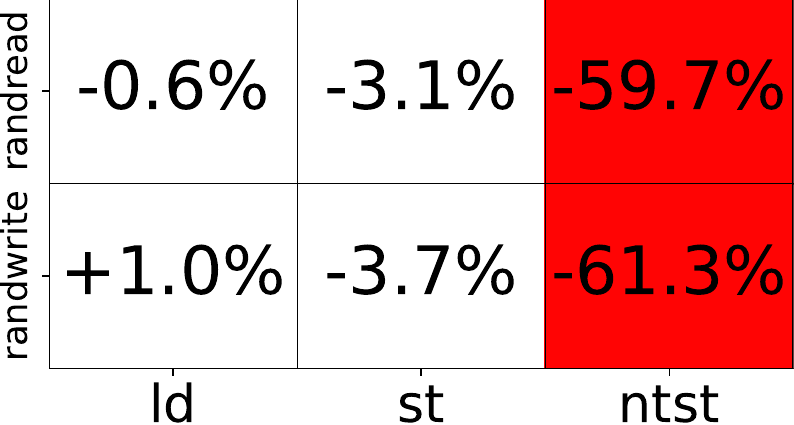} &
        \vspace{5pt}\includegraphics[width=0.95\linewidth,trim={0 0 0 0}]{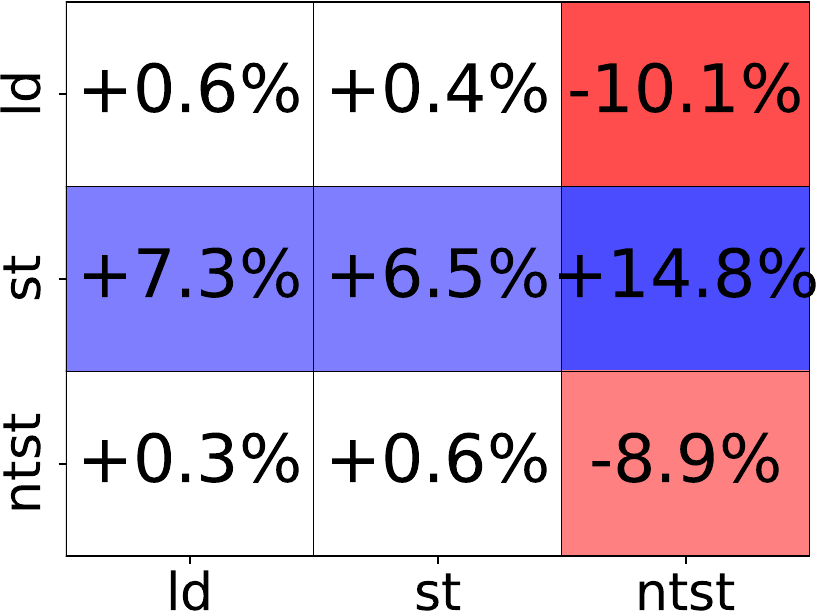} &        
        \vspace{5pt}\includegraphics[height=0.70\linewidth,trim={0 0 118 0},clip]{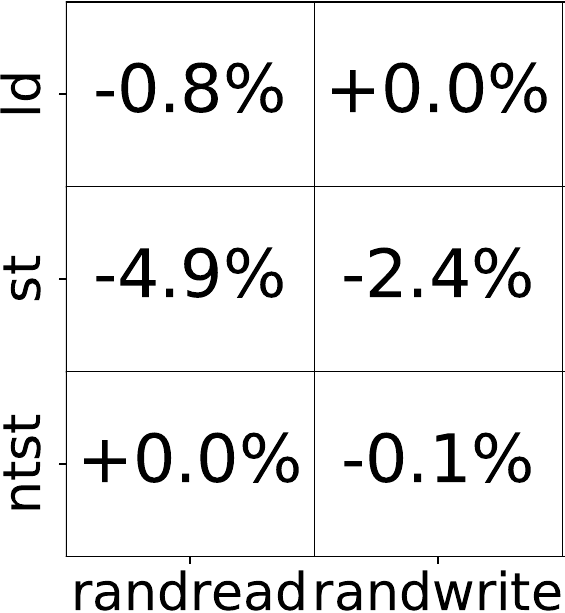} \\ \cline{2-6}
        & System B &    
        \vspace{5pt}\includegraphics[width=0.95\linewidth,trim={0 0 0 0},clip]{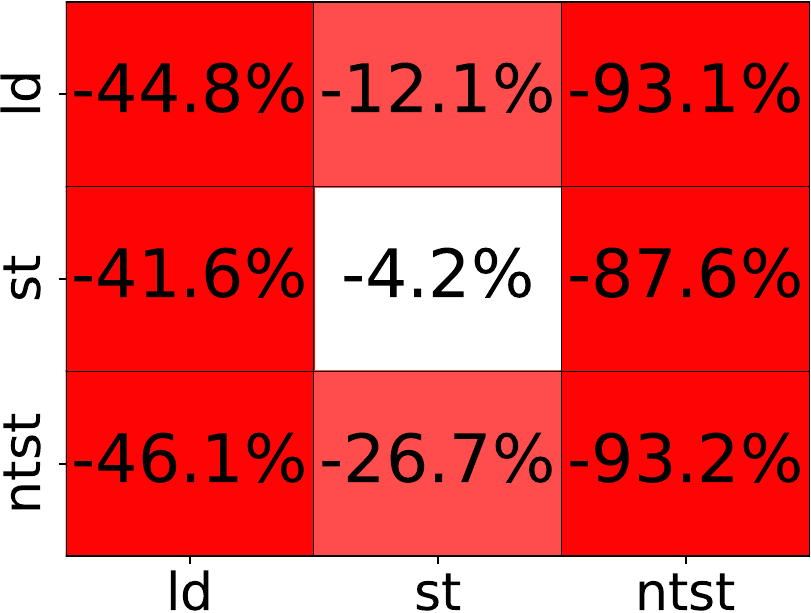} &
        \vspace{5pt}\includegraphics[width=0.95\linewidth,trim={0 0 0 90},clip]{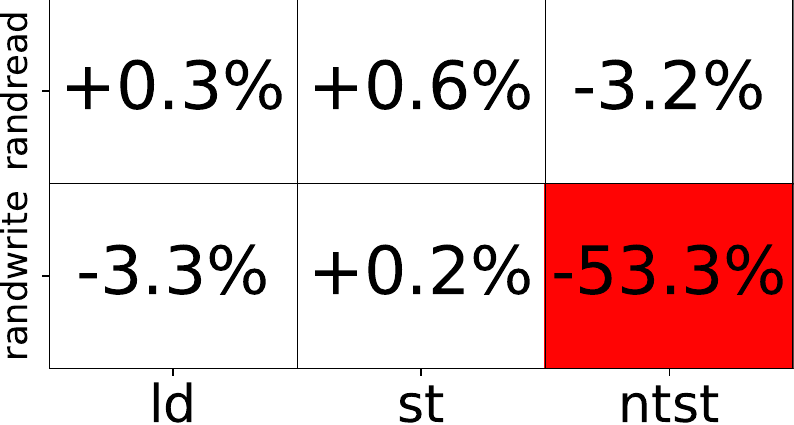} &
        \vspace{5pt}\includegraphics[width=0.95\linewidth,trim={0 0 0 0}]{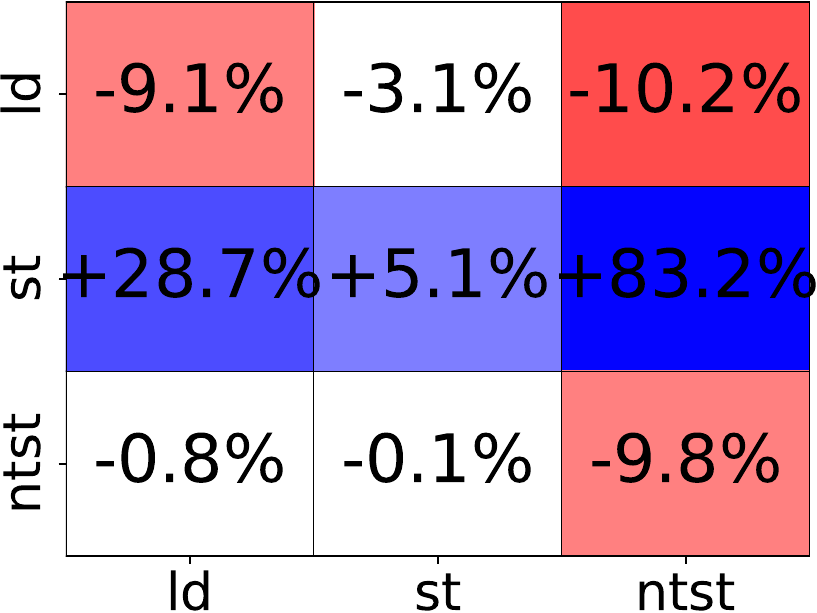} &
        \vspace{5pt}\includegraphics[height=0.70\linewidth,trim={0 0 118 0},clip]{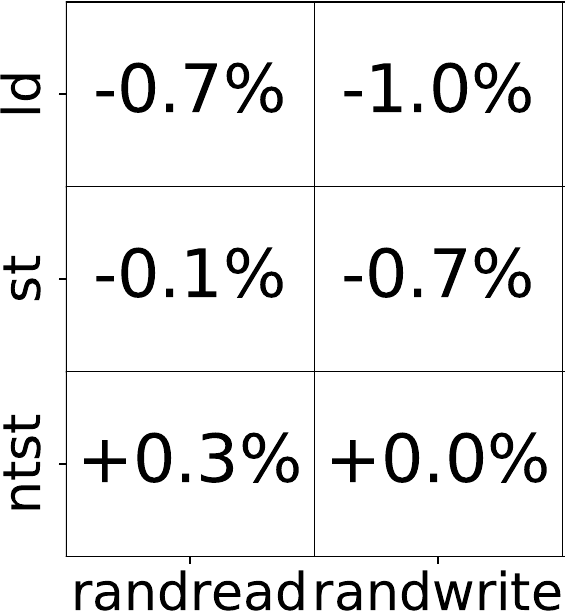} \\ \hline
        Colorbar & \multicolumn{5}{m{0.85\textwidth}|}{
        \centering
        \vspace{5pt}
        \includegraphics[width=0.95\linewidth]{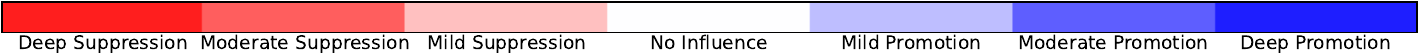}} \\
        \hline
    \end{tabular}

    \label{tab:mb}
\end{table*}

\subsection{Microbenchmark Characterization and Evaluation}
\label{subsec:micro-char}

\subsubsection{MMEM/SSD with CXL Traffic}

The left half of Table~\ref{tab:mb} illustrates the trends of MMEM/SSD's bandwidth under the interference of CXL. Each value represents the maximum interference observed across all thread count configurations. Our observations are as follows: a)   The bandwidth of MMEM/SSD suffers significant interference when CXL performs \textit{ntst}, reaching up to 93.2\%.  b) CXL \textit{ld} and \textit{st} can also be disruptive to MMEM, especially on System B, where over 40\% interference is observed. c) As for SSD, the interference from CXL \textit{ld} and \textit{st} traffic is relatively mild.

\noindent    \textbf{\textsc{INSIGHT \#1}: CXL traffic
(especially non-temporal store (\textit{ntst}))
  significantly impairs bandwidth of concurrent MMEM and SSD traffic, with MMEM experiencing greater degradation than SSD.} 

\subsubsection{CXL with MMEM/SSD Traffic}

The right half of Table~\ref{tab:mb} illustrates the bandwidth trends of CXL  under the interference of MMEM/SSD. Our observations are as follows: a) CXL bandwidth experiences performance boost  when CXL performs \textit{st} operations alongside MMEM traffic;
b) In other MMEM cases, \textit{ntst} operations exert a moderate to mild negative impact on CXL bandwidth.  
c) The bandwidth of CXL is not significantly influenced by SSD.

\subsection{Reverse-reasoning Analysis}
\label{subsubsec:Reverse-reasoning}

\begin{figure}[t!]
    \centering
    \includegraphics[height=0.5\linewidth]{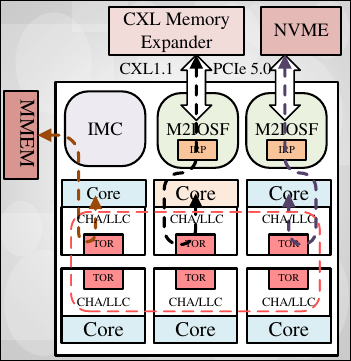}
    \caption{  {Datapaths in filesystem, CXL and MMEM workloads.} 
    }
    \label{fig:datapath}
\end{figure}

Figure~\ref{fig:datapath} illustrates the datapath for CXL memory, SSD and MMEM workloads. Figure~\ref{fig:datapath} depicts workload mixes involving SSD traffic and CXL traffic, where the CXL memory expander and storage device share the PCIe interface. 
It also shows workload mixes between   {CXL memory} and MMEM, where interfaces are not shared, but sharing of network-on-chip (NoC)  can  lead to interference.

\begin{table*}[t!]
    \centering
    \caption{Ratio of related event and function of NVMe.
    }
    \renewcommand{\arraystretch}{1.3}
    \begin{tabular}{>{\centering\arraybackslash}m{0.22\textwidth}>{\centering\raggedright}m{0.42\textwidth}>{\centering\arraybackslash}m{0.25\textwidth}}
        \hline
        \textbf{Related Event and Function} & \centering \textbf{Description} & \textbf{Interference to No-interference Ratio} \\ 
        \hline
        Load\_\L2\_Miss\_Latency & Average Latency for L2 cache miss demand Loads & 15.2$\times$ \\
        L3\_Cache\_Access\_BW & Average per-core data access bandwidth to the L3 cache & 0.14$\times$ \\
        Load\_L3\_Miss\_Latency & Average Latency for L3 cache miss demand Loads & 1.01$\times$ \\ \hline
        unc\_cha\_tor\_occupancy.ia\_miss & Occupancy number of TOR when request from iA misses LLC & 4090$\times$ \\ 
        unc\_cha\_tor\_inserts.ia\_miss & Inserts number of TOR when request from iA missses LLC & 155$\times$ \\ \hline
        memmove\_sse2\_unaligned\_erms() & Memory move with Enhanced REP MOVSB, used in memcpy() &1.96$\times$ \\ 
        copyout() & Kernel-to-User buffer copy  & 1.29$\times$ \\\hline
    \end{tabular}
    \label{tab:pmu}
\end{table*}

\textbf{Performance counters}. 
  {
We discover that when SSD is under interference of CXL, its L2 and local LLC miss average latency becomes much higher than the no interference scenario, which prompts an investigation into the PMU counters listed in Table{~\ref{tab:pmu}}. 
It can be observed that there is a significant increase in L2 miss latency, up to a factor of 15, accompanied by a notable reduction in bandwidth. However, the L3 miss latency remains relatively constant. This suggests that an examination of network-on-chip (NoC) between the L2 and L3 will be necessary. CPUs need to maintain a record of in-flight transactions that access the L3 slices that are attached to other cores. This record table presumably will, however, become a bottleneck.
}

  {
From the discovery and the results in Table{~\ref{tab:mb}} and the datapaths in Figure{~\ref{fig:datapath}}, we then focus on: a) Table of Request (TOR), a queue between cores and LLC, tracking pending Cache Home Agent (CHA) transactions{~\cite{SPR_events}}
of each core and PCIe interface; b) some OS \textit{kernel functions}.
}

Correspondingly, we collect performance monitor unit (PMU) listed in Table~\ref{tab:pmu}, where \(\frac{\textit{unc\_cha\_tor\_occupancy.ia\_miss}}{\textit{unc\_cha\_tor\_inserts.ia\_miss}}\) represents the average latency from cores that miss the LLC~\cite{SPR_events}.
We can deduce that data transferred by CXL can be propagated throughout the entire LLC, causing congestion in TOR due to TOR's high occupancy behavior and affecting memory accesses of other cores. This can be demonstrated from value change of \textit{unc\_cha\_tor\_inserts.ia\_miss} and \textit{unc\_cha\_tor\_occupancy.ia\_miss}. 

\noindent    {\textbf{\textsc{INSIGHT \#2}: CXL traffic occupies Table Of Request (TOR) for a lengthy period, causing congestion in TOR with other multicore process.}}

 \textbf{CXL counters}. We also evaluate CXL performance in both instruction level (with tools we used above) and in hardware level. We collect CXL-related PMU counters to compute CXL bandwidth. The results are consistent. For example, when CXL traffic with 16 \textit{ntst} threads run under the interference of SSD traffic with 16 \textit{randwrite} threads in System B, the bandwidth of CXL measured by cxlMemTest is 13169 MB/s, and the result computed by \textit{cxlcm\_txc\_pack\_buf\_inserts.mem\_data} is 13342 MB/s. The close agreement between the two values proves the reliability of our data.

\textbf{Kernel functions}. By profiling both SSD traffic alone and with CXL, two hotspot functions show significant changes. We notice that the time taken by \textit{memmove\_sse\_unaligned\_erms} and \textit{copyout} (kernel-to-user copy) increases by 96.2\% and 28.7\%, respectively, when CXL is in the background. These functions, called by \textit{memcpy} to move data in parallel, indicate that memory copy operations of system kernel processing data from devices are influenced.

\subsection{Microbenchmarks on Other Configurations}

\subsubsection{Sequential vs. Random Access} 
We also explore whether the interference differs when SSD performs sequential writes versus random writes. Results in Figure~\ref{fig:seq_rand} highlight   {that SSD is more susceptible to CXL interference when performing sequential writes (up to 80.9\% performance loss) compared to random writes (up to 54.5\% performance loss).} 
{The reason is attributed to the greater spatial locality of data in sequential writes. Random operations presumably encounter a large number of cache line conflicts and are limited by the small number of available DDIO ways. Consequently, random operations result in an increased number of cache misses and occupy a smaller proportion of LLC space (2MB) than sequential accesses (4MB), which makes sequential writes more vulnerable to interference from the cache side, originating from CXL.}

\noindent    {\textbf{\textsc{INSIGHT \#3}:  SSD sequential writes exhibit greater LLC occupancy, which are interfered more by CXL traffic than random writes.}}

\subsubsection{\texttt{MOVDIR64B} Operations}
\begin{figure}[t]
    \centering
    \begin{subfigure}{0.49\columnwidth}
        \centering
        \includegraphics[width=\linewidth]{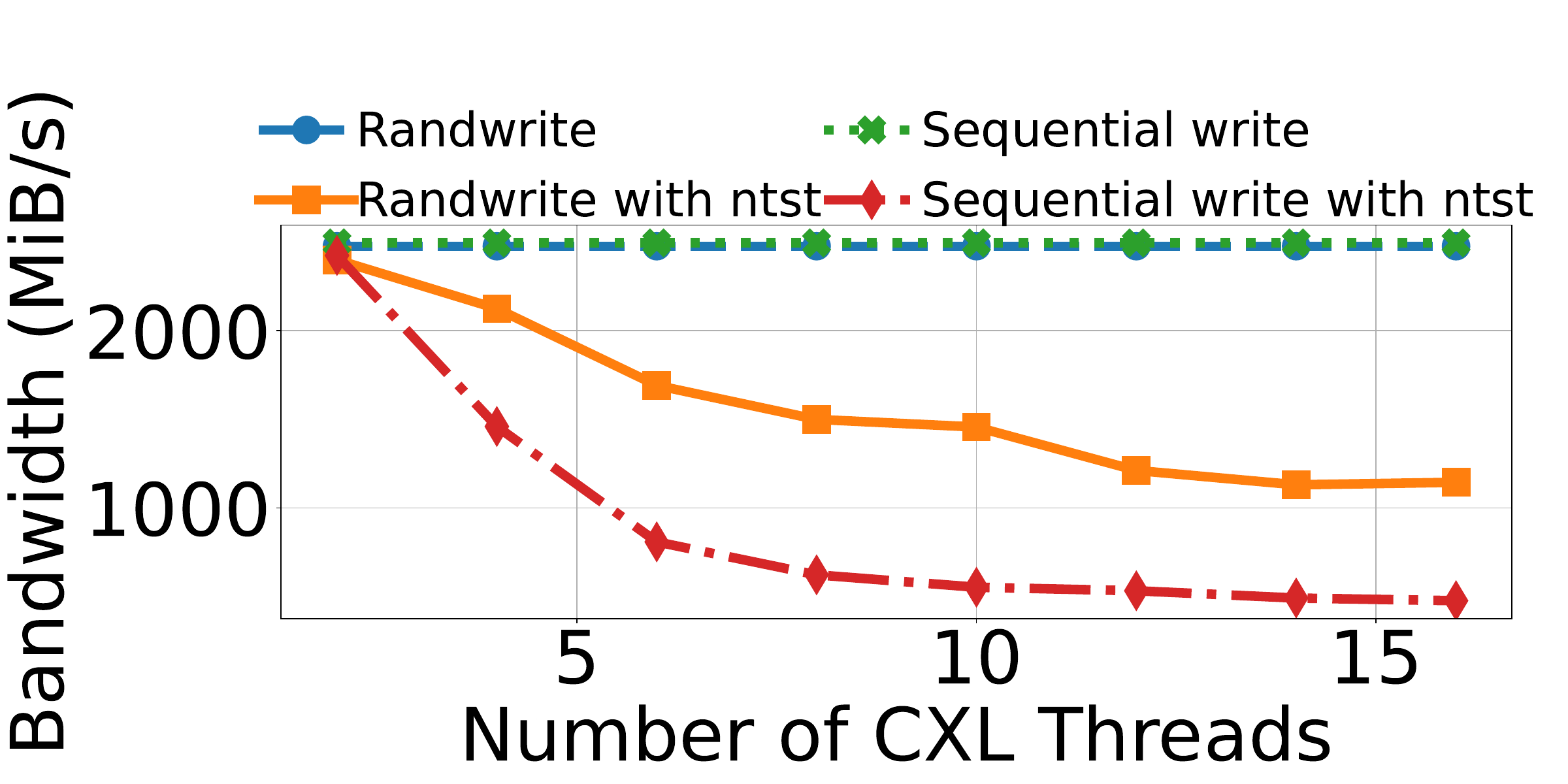}
        \caption{NVMe randwrite and sequential write bandwidth under the interference of CXL non-temporal store.}
        \label{fig:seq_rand}
    \end{subfigure}
    \begin{subfigure}{0.49\columnwidth}
        \centering
        \includegraphics[width=\linewidth]{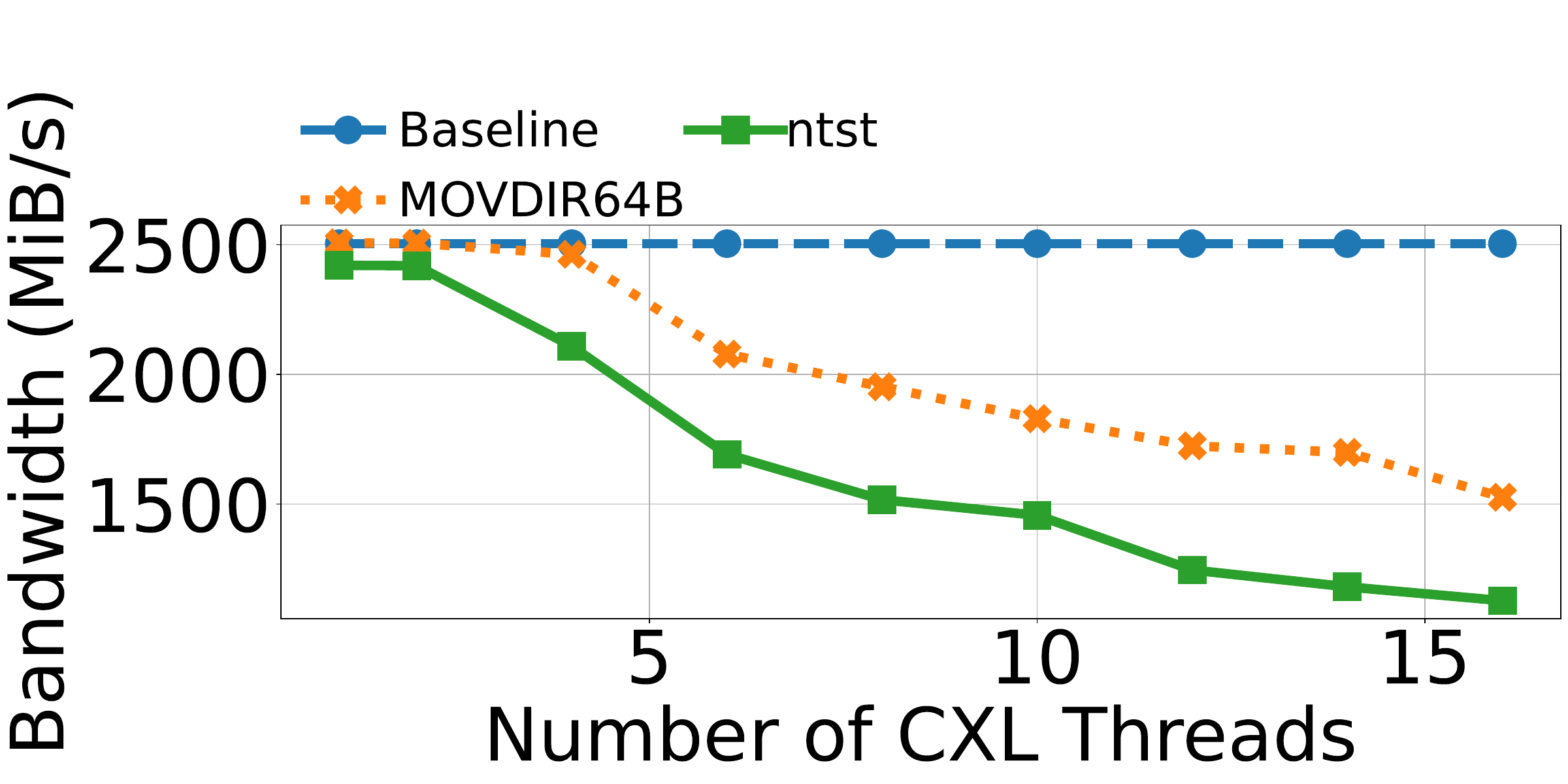}
        \caption{NVMe randwrite bandwidth across different memory access instructions.}
        \label{fig:movdir}
    \end{subfigure}
    \caption{Microbenchmark case studies}
\end{figure}

\texttt{MOVDIR64B}~\cite{intelsdm} is a specialized instruction that moves 64 bytes of data directly from source memory to destination memory with write atomicity. \footnote{This operation is available on the 4th generation Intel® Xeon® Scalable Processor Family based on Sapphire Rapids microarchitecture.} 
Compared to \textit{ntst}, \texttt{MOVDIR64B} adheres to the Write Combining (WC) memory type protocol and is weakly ordered. We wonder whether direct-store instructions like \texttt{MOVDIR64B} perform similarly to \textit{ntst}. 
Figure~\ref{fig:movdir} shows the influence of CXL \texttt{MOVDIR64B} on SSD \textit{randwrite} in comparison with  SSD performance alone and with CXL \textit{ntst}.  
{The results indicate that the interference from \texttt{MOVDIR64B} is similar to that of \textit{ntst}, 
but imposes relatively less suppression on SSD performance than \textit{ntst}. The presumable underlying cause is that unlike \textit{ntst}, \texttt{MOVDIR64B} is capable of immediate eviction from the WC buffer~\cite{intelsdm}. As a result, data of \texttt{MOVDIR64B} spends a shorter time within the WC buffer, reducing the cost of tracking outstanding instructions. Therefore, the impact of \texttt{MOVDIR64B} is less significant than that of \textit{ntst}. It is noteworthy that when the destination address of \texttt{MOVDIR64B} is local memory, thereby performing a normal load operation on CXL, the suppression disappears. This observation corroborates that the primary source of interference from CXL is its impact on the cache side.}

\noindent    {\textbf{\textsc{INSIGHT \#4}: CXL \texttt{MOVDIR64B} causes less SSD performance suppression than \textit{ntst} due to immediate WC buffer eviction, reducing instruction tracking costs.}}

%% file: 05_real_applications.tex
\section{\name{} Real Application Evaluations}
\label{sec:real_app}
\begin{figure*}[t]
    \centering
    \includegraphics[width=0.99\textwidth, trim={410 185 485 266}, clip]{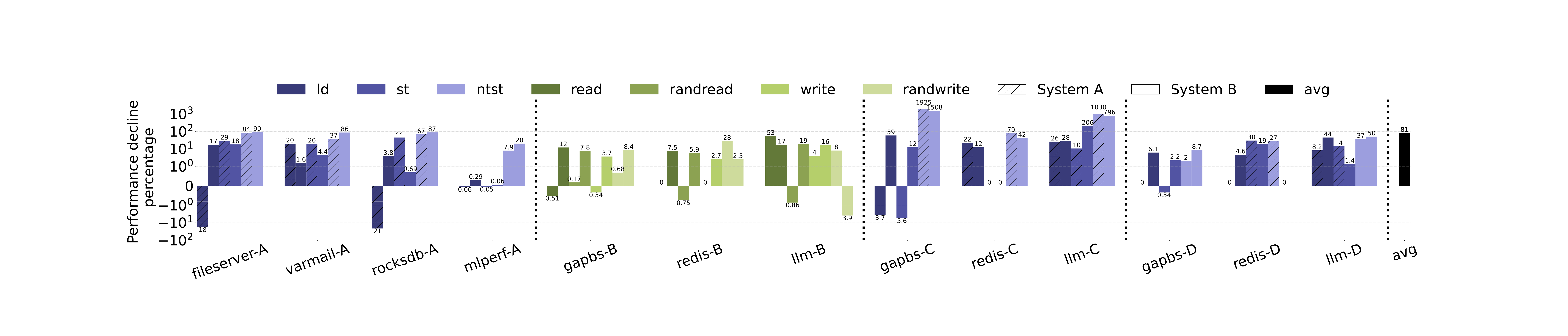}
    \caption{CXL-related interference on real applications. The suffixes "-A/B/C/D" indicate the type to which the experiment belongs. We use different colors to differentiate background traffic types. For Type A, C and D, the background runs memory traffic which have three different types: \textit{ld}, \textit{st} and \textit{ntst}. As for Type B, its background is file system and has four traffic types: \textit{read}, \textit{randread}, \textit{write} and \textit{randwrite}. We also use hatches to differentiate results from different systems. The height of the bars represents the maximum percentage of performance decline observed across all number of background thread count configurations for the corresponding scenarios. The y-axis is on logarithmic scale.
    \protect\footnotemark
    }
    \label{fig:barchart}
\end{figure*}

We select a wide range of applications with different characteristics to study the interference problem. There are 4 types of interference scenarios: 
filesystem-related applications under CXL traffic, CXL-related applications under SSD traffic, MMEM-related applications under CXL traffic and CXL-related applications under MMEM traffic. 
We denote them as Type A, Type B, Type C and Type D respectively. The description of the selected applications is shown in Table~\ref{tab:description}. The overview of the largest performance decline percentage evaluation result for each application is depicted in Figure~\ref{fig:barchart}. The key takeaway from this figure is that in most cases there is solid contention and interference across different access types and systems.

\begin{table}[t]
  \caption{Description of the selected applications.}
  \centering
  \renewcommand{\arraystretch}{1.3}
  \begin{tabular}{>{\centering\arraybackslash}m{0.12\columnwidth} | >{\centering\arraybackslash}m{0.17\columnwidth} | >{\centering\arraybackslash}m{0.55\columnwidth}}
  \hline
  \textbf{Category} & \textbf{Application} & \textbf{Description} \\ \hline
      \multirow{8}{0.12\columnwidth}{\centering File-related} &Filebench~\cite{filebench}: & An open-source file system benchmark \\ 
      & fileserver & Large files with 1:2 read/write ratio \\ 
      & varmail & Small files with 1:1 read/write ratio  \\ \cline{2-3}
      & \multirow{3}{0.17\columnwidth}{\centering RocksDB~\cite{rocksdb}} & Log-structured merge-tree database \\
      & & 20-byte key size and 100-byte value size \\
      & & \textbf{Metric:} Operations per second \\ \cline{2-3}
      & \multirow{2}{0.17\columnwidth}{\centering MLPerf Storage~\cite{balmau2022characterizing}} & Storage benchmark for deep learning  \\
      & & \textbf{Metric:} Accelerator Utilization \\ \hline
      \multirow{8}{0.12\columnwidth}{\centering Memory-related} & \multirow{3}{0.17\columnwidth}{\centering GAPBS~\cite{beamer2017gapbenchmarksuite}} & A graph computing benchmark \\
      & & PageRank Algorithm \\
      & & \textbf{Metric:} Trial time (lower is better)  \\ \cline{2-3}
      & \multirow{3}{0.17\columnwidth}{\centering Redis~\cite{redis}} & An open-source in-memory database \\ 
      & & 50\% read and 50\% update \\
      & & \textbf{Metric:} Operations per second   \\ \cline{2-3}
      & \multirow{2}{0.17\columnwidth}{\centering Large Language Model} & TinyLlama~\cite{zhang2024tinyllama}, a decoder-only LLM \\[0.3em]
      & & \textbf{Metric:} Inference time (lower is better)\\ \hline
  \end{tabular}
  \label{tab:description}
\end{table}

\subsection{Filesystem-Related Applications}

We select RocksDB database, Filebench, and machine learning workloads MLPerf as our representative applications, 
while using CXL traffic as background to measure the interference for filesystem.

\begin{figure}[t]
    \centering
    \begin{subfigure}{0.49\columnwidth}
        \centering
        \includegraphics[width=\linewidth]{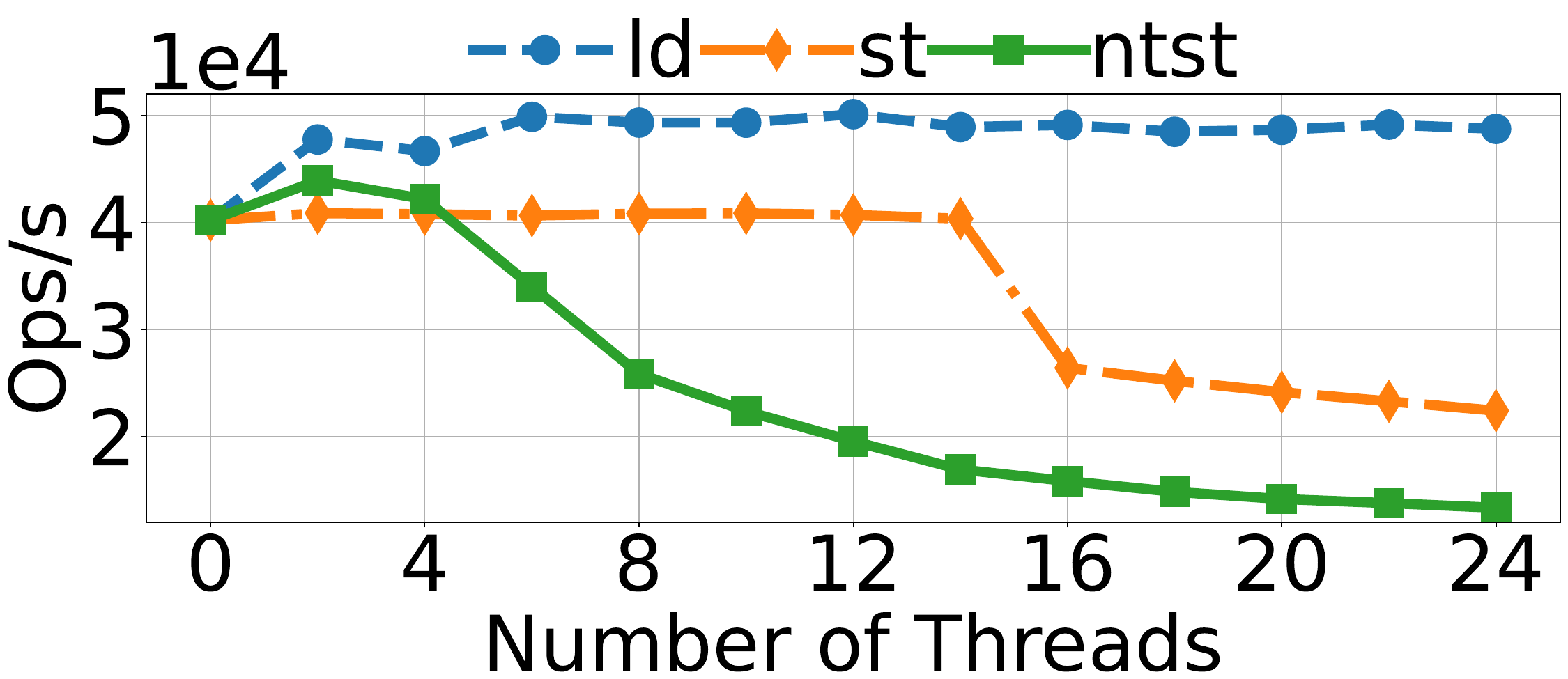}
        \caption{System A}
        \label{fig:rocks_sas}
    \end{subfigure}
    \begin{subfigure}{0.49\columnwidth}
        \centering
        \includegraphics[width=\linewidth]{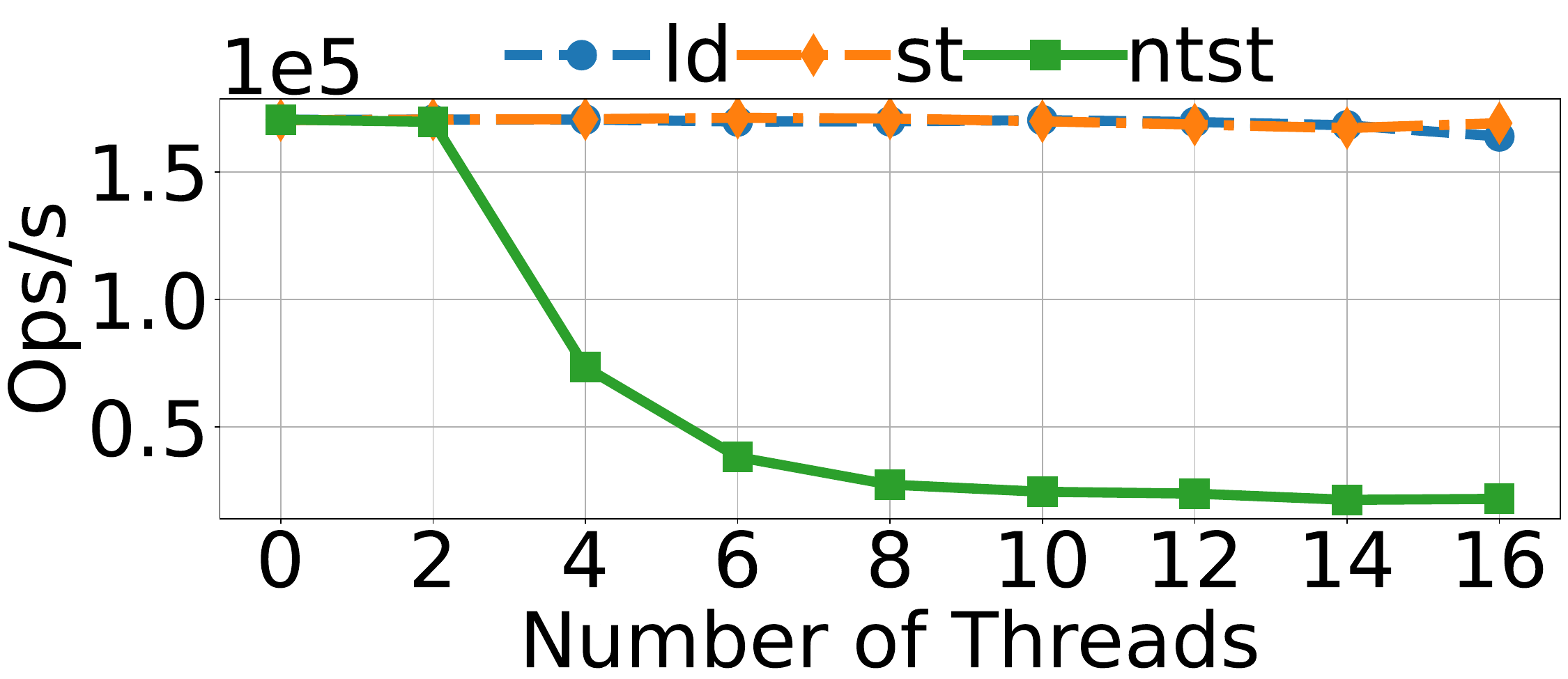}
        \caption{System B}
        \label{fig:rocks_ssd}
    \end{subfigure}
    \caption{The performance (operations per second) of RocksDB under CXL traffic. The x-axis indicates the number of background threads.}
    \label{fig:rocks}
\end{figure}

Figure~\ref{fig:barchart} demonstrates that CXL's traffic has a considerable effect on the performance of most applications (Type A), with an observed detrimental impact reaching up to 90\% for \textit{ntst}. For \textit{ld} and \textit{st}, the impact can also be as high as 20\% and 44\% respectively. 
Moreover, the results presented in Figure~\ref{fig:rocks} indicate an interesting and unexpected phenomenon that CXL’s ld traffic can even enhance RocksDB’s performance by up to 21.2\% on system A.

\footnotetext{In certain scenarios, the results are unstable and exhibit significant fluctuations. In such cases, the maximum decline is not meaningful, and we represent it as 0 in the figure.}

\textbf{Discussion.} It can be observed that an increase in the thread count—and consequently the bandwidth—of the background CXL \textit{ld} workload is associated with improved performance. To verify this, two other methods are applied to adjust CXL memory bandwidth: periodic cache flushing and injecting numerous \texttt{nop} instructions. We observe that the performance enhancement is directly linked to CXL memory bandwidth, occurring only when it exceeds 8 GB/s. Between 8 GB/s and 20 GB/s, the promotion grows linearly, while beyond that, the enhancement plateaus.

Albeit system-specific, we try to explore the root cause for promotion in System A with performance counters
for RocksDB, 
running a fixed number of operations both with and without CXL \textit{ld} interference. The most notable observation is a 24\% reduction in the number of instructions, especially memory instructions, under interference. This reduction in instructions plays a key role in the performance improvement.

\noindent    {\textbf{\textsc{INSIGHT \#5}: SSD workload bandwidth can benefit from a reduced instruction count when having sufficient CXL \textit{ld} memory bandwidth as background.}}

\subsection{Memory-Related Applications}

\begin{figure}[t]
    \centering
    \begin{subfigure}{0.49\columnwidth}
        \centering
        \includegraphics[width=\linewidth]{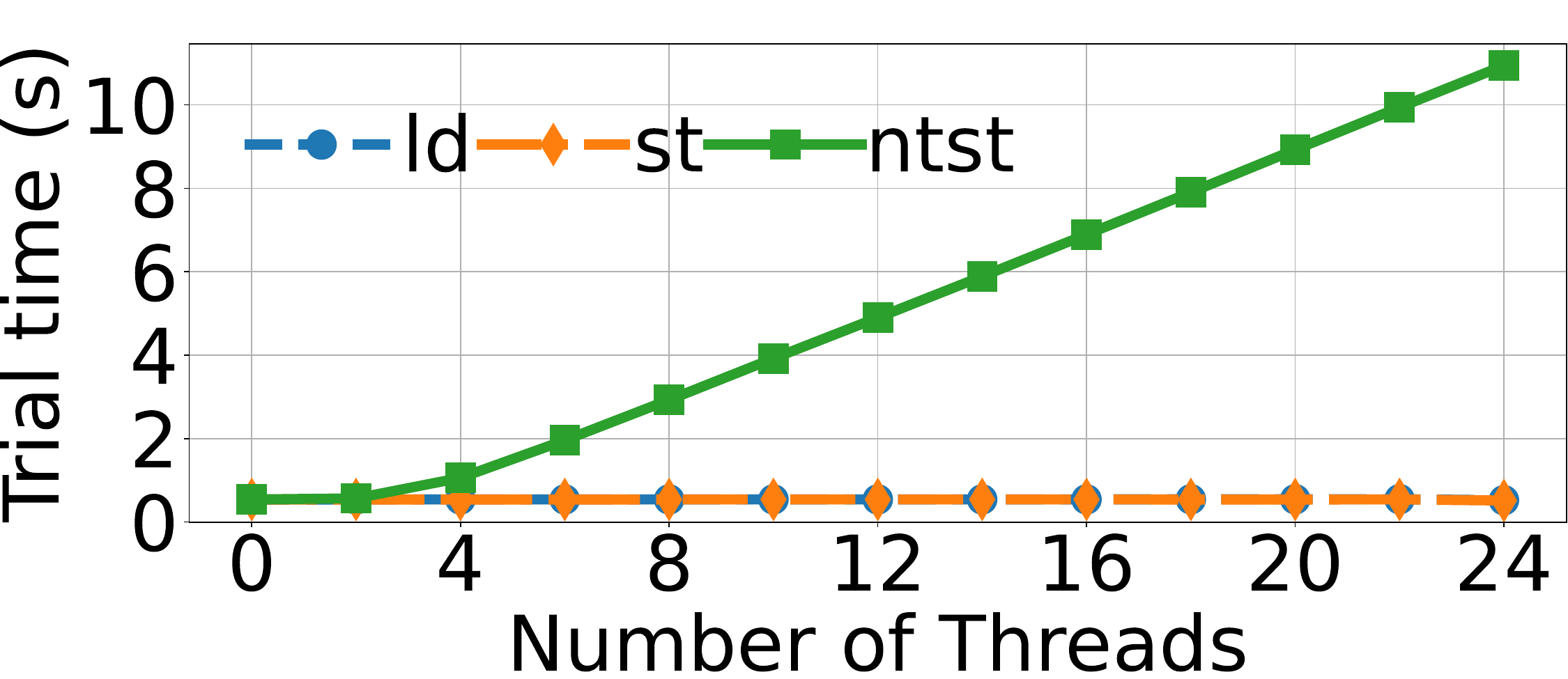}
        \caption{System A}
        \label{fig:gapbs_sas_typeC}
    \end{subfigure}
    \begin{subfigure}{0.49\columnwidth}
        \centering
        \includegraphics[width=\linewidth]{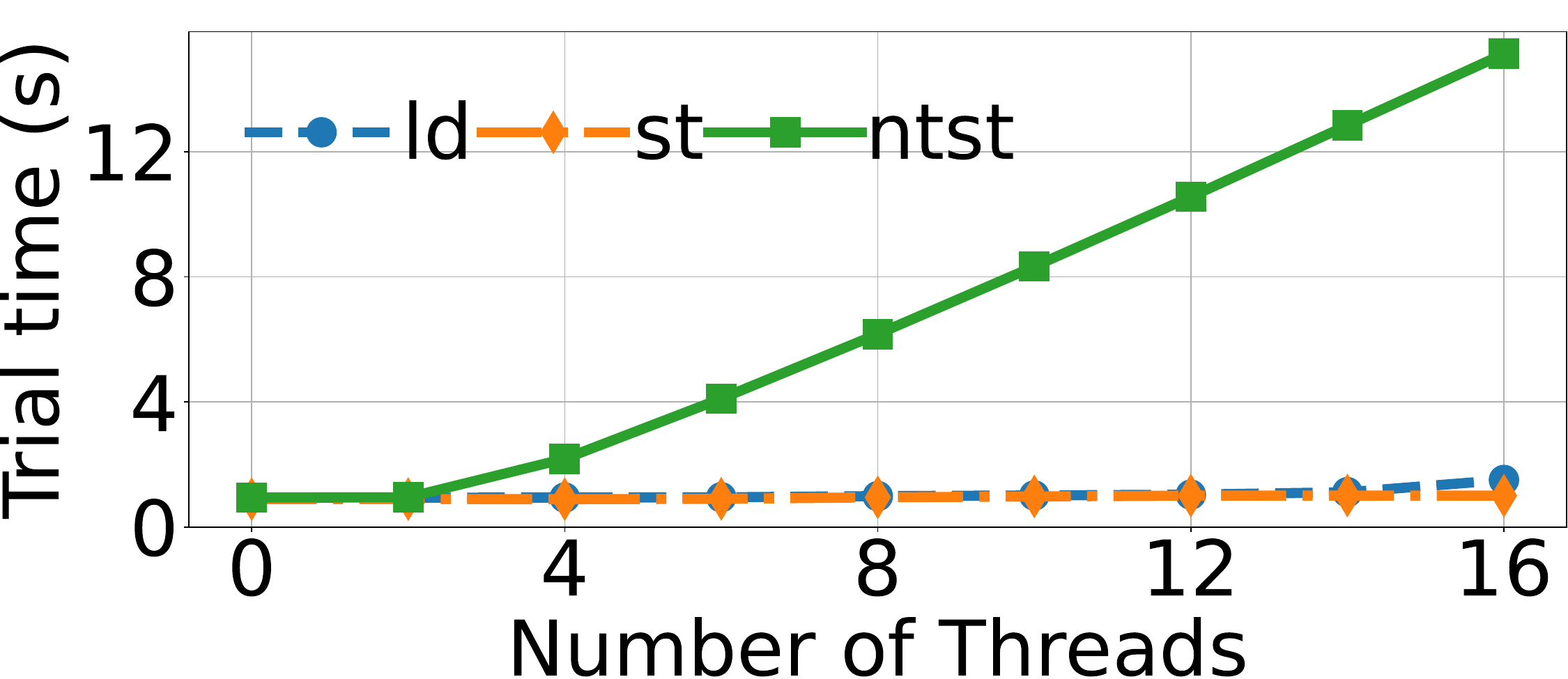}
        \caption{System B}
        \label{fig:gapbs_ssd_typeC}
    \end{subfigure}
    \caption{  {The performance (trial time) of GAPBS benchmark in Type C. The x-axis indicates the number of background threads. }}
    \label{fig:gapbs_typeB}
\end{figure}

We select graph benchmark GAPBS, in-memory database Redis and an LLM as our representative memory applications, covering Type B, C and D. 
It can be clearly observed that CXL memory's operations have a substantial impact on MMEM (Type C), consistent with microbenchmark results. 
Figure~\ref{fig:gapbs_typeB} shows a representative with up to 20.8$\times$ performance decline.

Conversely, the impact of MMEM on CXL memory is significantly less pronounced than that observed in Type C. We hypothesize that the observed limited impact is due to the fact that, although MMEM traffic is high, the low latency of MMEM does not result in significant consistence overhead, allowing CXL to operate normally.

Regarding SSD applications, their impact on CXL memory is insignificant, potentially due to the fact that the LLC consumed by SSD is a relatively small section (4MB) in comparison to those occupied by MMEM (50MB). Nevertheless, it should be noted that in some cases, particularly under \textit{read} operation background, up to 12\% decline in CXL performance can be observed.

\noindent    {\textbf{\textsc{INSIGHT \#6}: The interference from CXL memory to MMEM is significant. However, MMEM and SSD have mild impact on CXL memory in most cases due to DRAM's lower latency and SSD's lower LLC utilization. 
}
} 

%% file: 06_hardware_influence_on_qos.tex
\section{Intervent the interference and Experiment on CXL-Related Regulation}
\label{sec:hardware}

Given the interference discussed in Section~\ref{sec:microbenchmarks} and Section~\ref{sec:real_app}, we have observed and analyzed various interplay between CXL and SSD, as well as CXL and MMEM. In this section, we provide different trials and solutions to mitigate the interference with real CXL hardware.
We consider using CPU usage restriction, such as  time quota allocating, frequency scaling, memory bandwidth restriction, etc., to demonstrate the effectiveness of regulation on CXL real hardware.

\subsection{CPU Usage Restrictions}
\label{subsec:usage_restrict}

  {To validate the functionality of CPU usage restriction, we choose three scenario of severe interference: MMEM \textit{ld}, MMEM GAPBS and filebench \textit{varmail}, all influenced by 16 CXL \textit{ntst} threads.}
We first use the counters mentioned in Section~\ref{subsubsec:Reverse-reasoning} and bandwidth monitoring tools~\cite{pcm} to identify the processes causing high CXL traffic and LLC occupancy periodically. Then we use Linux's \textit{cgroups}~\cite{cgroups-v2} to limit their CPU quota. If the process still causes severe impact in the next period, we further adjust its resource allocation to achieve dynamic control.

\begin{figure}[t]
    \centering
    \begin{subfigure}{0.49\columnwidth}
        \centering
        \includegraphics[width=\linewidth]{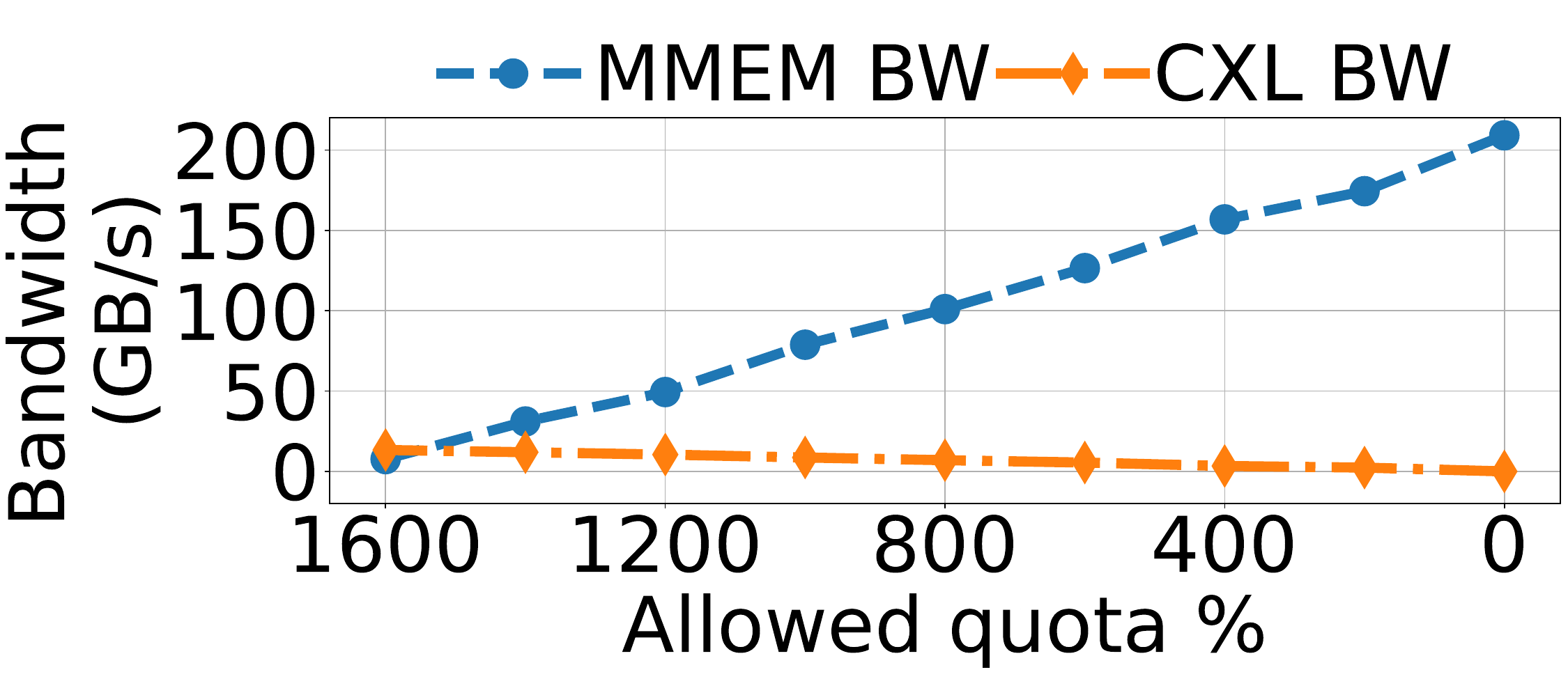}
    \end{subfigure}
    \begin{subfigure}{0.49\columnwidth}
        \centering
        \includegraphics[width=\linewidth]{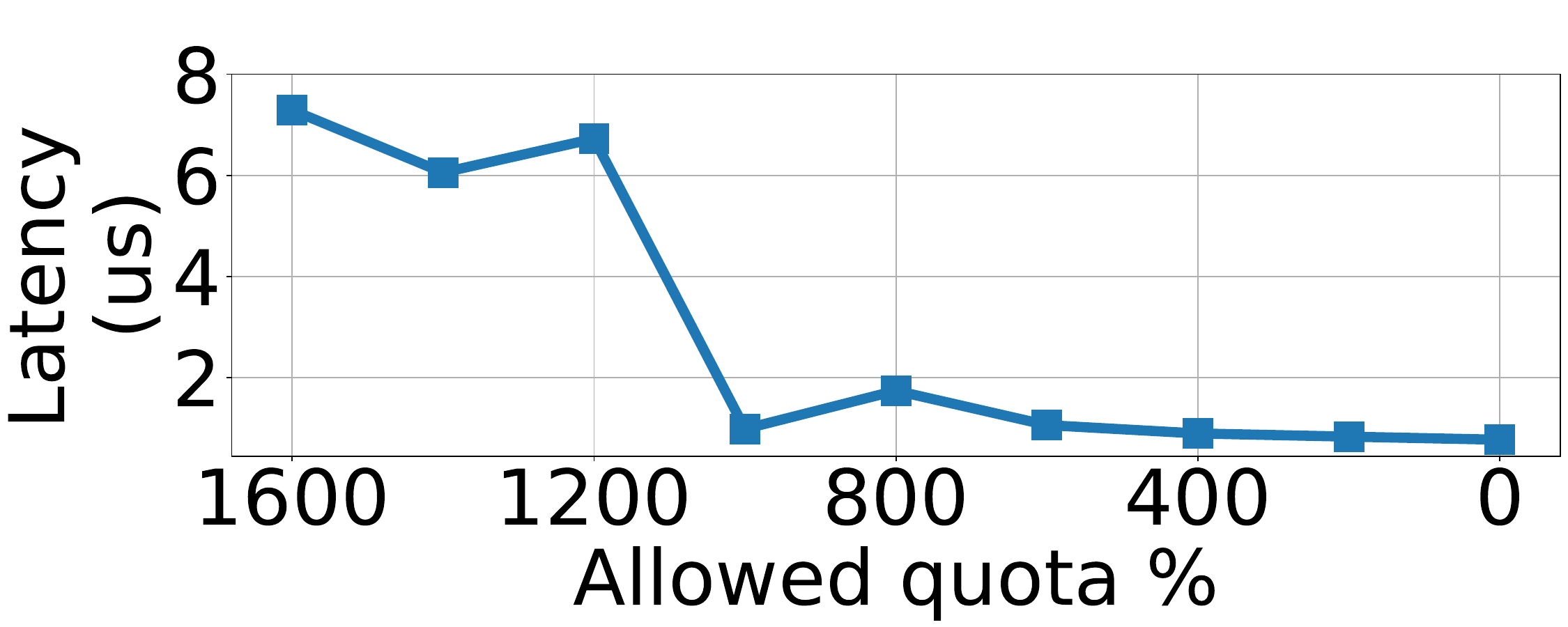}
    \end{subfigure}
    \caption{  {Bandwidth and latency of MMEM when applying different CPU usage quotas to CXL threads. The horizontal axis is the allowed percentage of CPU usage.}}
    \label{fig:qos_soft_dram}
\end{figure}

As illustrated in Figure~\ref{fig:qos_soft_dram}, reducing the allowed CPU usage quota leads to a recovery in MMEM bandwidth and a decrease in latency. Specifically, when the distribution of CPU cycles for CXL is restricted to 100\% ($1/16$ of the time originally occupied), the MMEM bandwidth recovers to   {$190$ GB/s, which is approximately 92\% of the bandwidth} observed without interference. Additionally, latency returns to nearly its original level. This improvement in MMEM performance comes at the cost of a bandwidth reduction of about 10 GB/s on CXL.
The results for the three applications is illustrated in Figure{~\ref{fig:qos_soft}}.
The linear and consistent impact observed across the three applications are due to the characteristic of CPU usage regulation, which influences a wide range of instructions. This approach facilitates a satisfactory and adjustable degree of recovery, albeit with varying degrees of performance decline in the background process.

\subsection{Frequency Scaling}
\begin{figure}[t]
    \centering
    \begin{subfigure}{0.49\columnwidth}
        \centering
        \includegraphics[width=\linewidth]{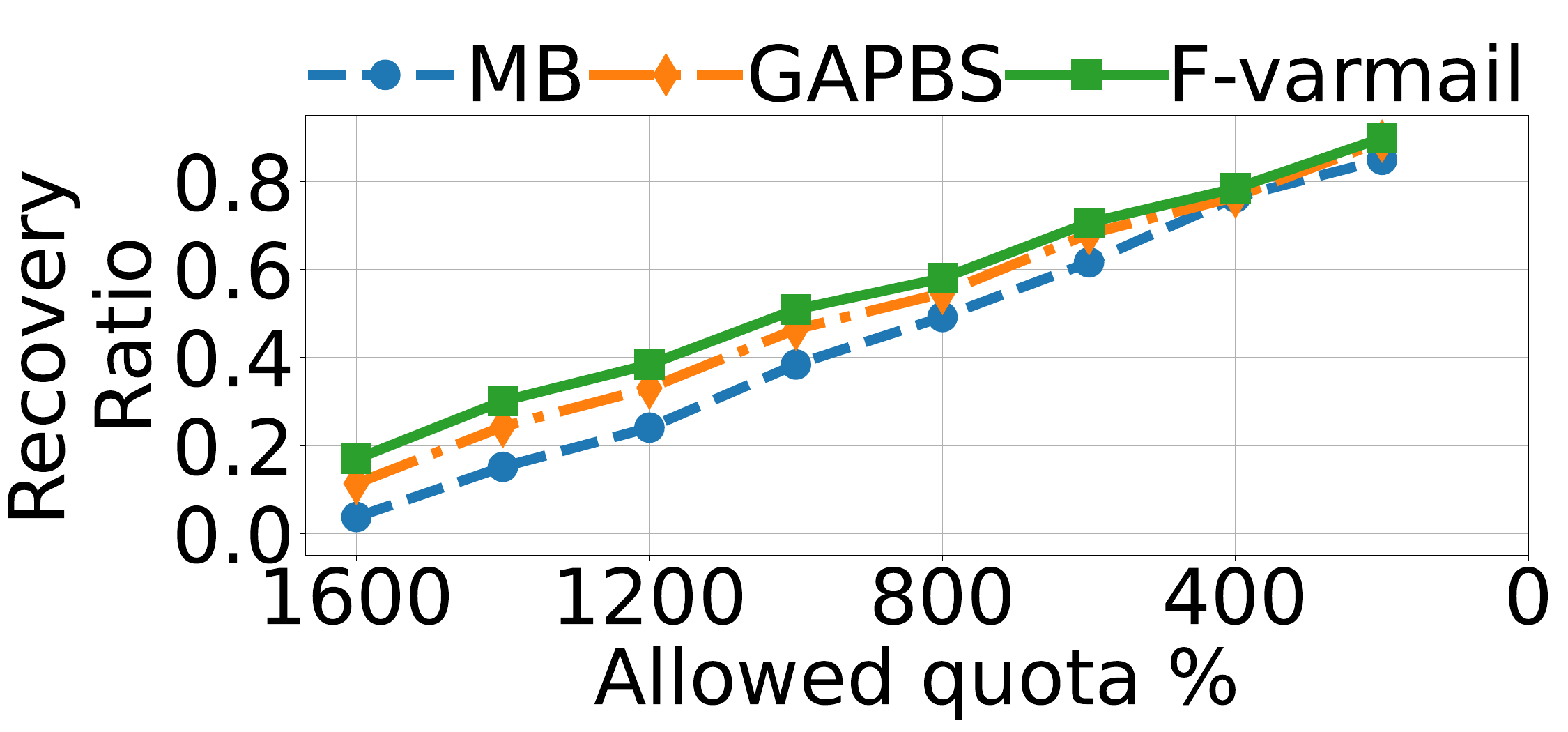}
        \caption{CPU usage restriction}
        \label{fig:qos_soft}
    \end{subfigure}
    \begin{subfigure}{0.49\columnwidth}
        \centering
        \includegraphics[width=\linewidth]{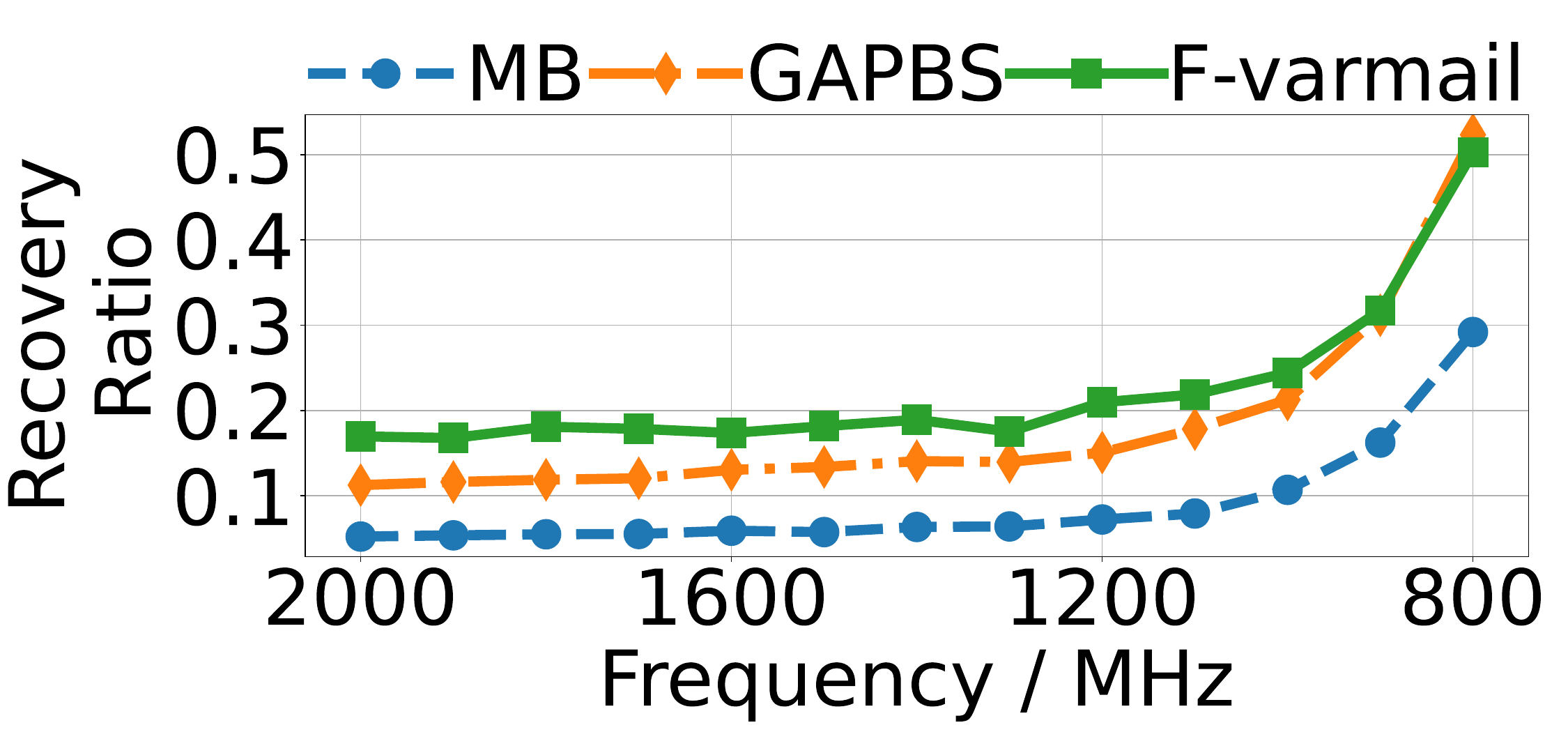}
        \caption{Frequency Scaling}
        \label{fig:qos_freq}
    \end{subfigure}
    \caption{Effectiveness of the two methods. Vertical axis is normalized to the performance without interference.}
    \label{fig:qos_soft_freq}
\end{figure}

We also consider frequency scaling as a potential solution to mitigate the interference between CXL and MMEM. As is shown in Figure{~\ref{fig:qos_freq}}, the recovery of MMEM threads is observed to be within the range of 30\% to 52\% when the background is constrained to very low frequency.
This phenomenon can be explained by the fact that, except for very low frequency cases, memory instructions continue to fill the load/store queue. Consequently, memory-intensive tasks tend to demonstrate an exponential effect as the frequency decreases, yet still exhibit a suboptimal recovery rate.
Although this method is less effective, it does not cause any damage to the memory-only background program.

\subsection{Memory Bandwidth Restriction}
The solution of regulating time slice and frequency is purely software based and universal. But it also restricts the execution of memory-unrelated instructions, which causes additional performance degradation. 
On the other hand, memory bandwidth restriction presents a promising approach to mitigating interference between CXL and MMEM.
We evaluate the effectiveness of using Intel MBA{~\cite{intelrdt}}, which is a feature that allows users to allocate a portion of the memory bandwidth to specific processes.
Limited by the I/O block configuration, it is only possible to restrict the CXL bandwidth to approximately less than half of its original value, which corresponds to a reduction of 5GB/s.
This approach proves highly effective, enabling the three applications to recover to 99\%, 95\%, and 98\% respectively, without affecting the compute instructions.
However, this method is unable to provide control in a responsive manner and takes seconds to take effect, making it more suitable for implementation on the CXL device side.

\subsection{Discussions on Other Methods}
There are multiple solutions proposed to restrict the noisy neighbors and mitigate their negative impact on other threads.   {
Apart from the methods discussed above,
other methods include 1) cache partitioning, which is enabled by CPU vendors, such as Intel's Cache Allocation Technology (CAT) and 2) specialized memory bandwidth restriction, such as AMD's Slow Memory Bandwidth Allocation (SMBA){~\cite{amd64qos}} targeting CXL devices, which has not been released.
However, cache partitioning has minimal impact on operations that bypass the cache hierarchy, such as \textit{ntst} operations, where it proves ineffective.
}

For hardware implementation, additional latency can be inserted to CXL device side after CPU detecting actual interference and then write the control register to change the added latency. However, this modification can highly depend on specific hardware which makes it harder to quickly deploy and generalize to different vendors.

%% file: 05p_practical_suggestion.tex
\section{Practical Takeaways}

  {
 Given the characterization and regulation experiments of CXL interference above, we provide the following takeaways for software (S) and hardware (H) developers respectively when developing future CXL-related systems:}

 \begin{enumerate}
     \item[  {(S1)}]   {Limit the number of threads or bandwidth of CXL \textit{ntst} operations when possible to minimize their impact on MMEM and SSD workloads. (\#1)}
     \item[  {(S2)}]   When running under CXL \textit{ntst} traffic, SSD workloads should consider the trade-off between different access patterns such as random vs. sequential writes to maintain bandwidth. (\#3) 
     \item[  {(S3)}]   {For some systems, it is beneficial of lifting bandwidth to schedule SSD workloads with  workloads having intensive CXL memory \textit{ld} operations. (\#5)}
 \end{enumerate}
 \begin{enumerate}
     \item[  {(H1)}]   {Hardware vendors can develop specialized TOR for slow tier memory to mitigate CXL contention with MMEM and SSD workloads. (\#2)}
     \item[  {(H2)}]   {To regulate the interference, it is advised to monitor and regulate on the CXL.mem device side  with a more fine-grained MBA-like regulator.}
 \end{enumerate}

%% file: 08_conclusion.tex
\section{Conclusion}
\label{sec:conclusion}

Real devices utilizing the emerging CXL technology are now finally commercially available. Researchers have actively characterized their features as isolated components. In this paper, we propose that since CXL devices need to interact with other components, it is necessary to characterize them in an coexisting manner. Our evaluation shows that this interference can cause a performance drop of up to 93.2\%, along with other interesting observations.
We also delve deep into the root causes of these observations. To mitigate the impact of this issue, we propose mechanisms to manage CXL traffic. The results demonstrate the efficacy and trade-off of several methods.  
\clearpage